# How biomedical papers accumulated their clinical citations: A large-scale retrospective analysis based on PubMed

Xin Li (0000-0002-8169-6059) xl60@hust.edu.cn
School of Medicine and Health Management, Tongji Medical College, Huazhong University of Science and Technology, Wuhan 430030, China

Xuli Tang* (0000-0002-1656-3014) xltang@ccnu.edu.cn
School of Information Management, Central China Normal University, Wuhan 430079, China

Wei Lu (0000-0002-0929-7416) weilu@whu.edu.cn
School of Information Management, Wuhan University, Wuhan 430074, China

**Abstract**

This paper explored the temporal characteristics of clinical citations of biomedical papers, including how long it takes to receive its first clinical citation (the initial stage) and how long it takes to receive two or more clinical citations after its first clinical citation (the build-up stage). Over 23 million biomedical papers in PubMed between 1940 and 2013 and their clinical citations are used as the research data. We divide these biomedical papers into three groups and four categories from clinical citation level and translational science perspectives. We compare the temporal characteristics of biomedical papers of different groups or categories. From the perspective of clinical citation level, the results show that highly clinically cited papers had obvious advantages of receiving clinical citations over medium and lowly clinically cited papers in both the initial and build-up stages. Meanwhile, as the number of clinical citations increased in the build-up stage, the difference in the length of time to receive the corresponding number of clinical citations among the three groups of biomedical papers significantly increased. From the perspective of translational science, the results reveal that biomedical papers closer to clinical science more easily receive clinical citations than papers closer to basic science in both the initial and build-up stages. Moreover, we found that highly clinically cited papers had the desperate advantage of receiving clinical citations over even the clinical guidelines or clinical trials. The robustness analysis of the two aspects demonstrates the reliability of our results. The indicators proposed in this paper could be useful for pharmaceutical companies and government policy-makers to monitor the translational progress of biomedical research. Besides, the findings in this study could be interesting for young scholars in biomedicine to get more attention from clinical science and to obtain success in their scientific careers, especially for those in basic science.

**Keyword**
Clinical citation; Temporal characteristics; Clinical citation level; Translational science; Robustness analysis

# 1 Introduction

The clinical impact of biomedical research has become increasingly important to academic activities in biomedicine, such as clinical translation, drug discovery, and grant application (Caulley et al., 2020; Li et al., 2020; Spencer, 2022). The government, pharmaceutical companies, and the people are more interested in biomedical research's clinical impact on illness treatment and health promotion than their scientific impact (Grant et al., 2000; Li et al., 2022). Clinical citation, which refers to the citations of biomedical papers received from clinical guidelines or clinical trials, has been a valuable measure for characterizing the clinical impact of biomedical research. Biomedical papers with more clinical citations were considered more accessible to be translated clinically (Hutchins et al., 2019; Thelwall & Kousha, 2016). However, it is noteworthy that many biomedical papers have never been clinically cited (Li et al., 2022). Therefore, a deeper understanding of the temporal characterization of the whole process of biomedical papers receiving their clinical citations will be helpful for us to accelerate the clinical translation of biomedical research.

The temporal features of a scientific paper receiving citations have been widely studied in bibliometrics, especially for the time required to receive its first citation (i.e., from 0 to 1). Egghe (2000) considered that a paper that received its first citation in a shorter time could have higher quality since the first citation represented the status of the paper transformed from "unseen" to "used". Rousseau (1994) and Wallace et al. (2009) pointed out that a paper might be cited for the first time shortly after its publication, if possible. Mathematically, Rousseau (1994), Egghe (2000), Burrel (2001), and Egghe & Ravichandra Rao (2001) presented their first-citation distributions based on various mathematical models, such as double exponential and Poisson models. Meanwhile, a series of quantitative indicators were proposed to measure the time to receive the first citation, for example, the Immediacy Index, the First-Citation-Speed-Index (Egghe et al., 2011), and the Response Time (Huang et al., 2019; Schubert & Glänzel, 1986). Particularly, when a paper took a long time to receive its first citation and became highly cited in a very short period afterward, it was delayed recognized and called a "sleeping beauty". Indicators such as the Beauty Coefficient (Ke et al., 2015), the Gs index (Min et al., 2016), and the Bcp (Du & Wu, 2018) have also been proposed to recognize sleep beauties in literature and patents. These studies provided us with a quantitative understanding of the first citation. However, they did not discuss the first clinical citation of a paper or compare the difference in the process of receiving the first clinical citation between various categories of biomedical papers, such as basic papers and clinical papers.

Citation distribution over time has also received a great deal of attention in the field of bibliometrics. Papers that received their first citations were more likely to accumulate more citations afterward (i.e., from 1 to N). Most of the extant related research did not distinguish between these two stages but focused on the whole process of citation (i.e., from 0 to N). For example, Radicchi et al. (2008) and Golosovsky (2021) claimed that the citation distributions are almost identical across different disciplines in science. Wallace et al. (2009) proposed a random selection method to analyze the dynamics of the citation distribution of 25 million papers published between 1900-2006. Waltman et al. (2012) concluded that the universal citation distribution across disciplines was not warranted based on a large-scale validation study. Features of citation distributions such as breadth and depth (Yang & Han, 2015), inequality (Nielsen & Andersen, 2021), network features (Goldberg et al., 2015), and citation level (Huang et al., 2021) were also profoundly examined. Huang et al. (2019) first compared the citation distributions of CS papers between the two stages, i.e., the beginning stage (from 0 to 1) and the accumulative stage (from 1 to N), and they also designed an indicator called "incremental time" to measure the "difficulty" in receiving citations.

Nevertheless, none of the existing studies have discussed the distribution of clinical citations of biomedical papers, let alone their temporal characteristics. Therefore, understanding how biomedical papers (especially basic papers) accumulate their clinical citations from the perspectives of clinical citation level and translational science was the focus of the current study. First, it shed light on the time characteristics of the process of biomedical papers being clinically cited that reflected the temporal process of clinical translation. According to our analyses, basic papers averagely needed to accumulate approximately 47% citation counts or 25 citation counts in their citation sequences before they received their first clinical citations, which were considered a sign of clinical translation. Second, we identified temporal characteristics of biomedical papers accumulating their clinical citations from two perspectives: clinical citation level and translational science. From the clinical citation level perspective, it was found that highly clinically cited papers had obvious advantages of receiving clinical citations over medium and lowly clinically cited papers in the whole process (i.e., from 0 to N). In the build-up stage (i.e., from 1 to N), as the number of clinical citations increased, the differences in the length of time to receive the corresponding number of clinical citations among the three groups of biomedical papers significantly increased. From the translational science perspective, the results showed that biomedical papers closer to clinical science more easily receive clinical citations than papers closer to basic science. Besides, the results also revealed that highly clinically cited papers had the desperate advantage of receiving clinical citations over even the clinical guidelines or clinical trials. Third, it was beneficial to identify differences in accumulating clinical citations among different categories, and different time phases (before or after being clinically cited for the first time) of biomedical papers, and thus help research managers to make scientific policies to promote the clinical translation of biomedical research.

## 2 Related work on clinical citations of biomedical papers

The number of clinical citations of biomedical papers has increasingly been known as a potential indicator for quantifying the clinical impact or clinically translational progression of biomedical projects, papers, or authors (Caulley et al., 2020; Kane et al., 2022; Kryl et al., 2012; Lewison & Sullivan, 2008; Spencer, 2022; Thelwall & Kousha, 2016). Despite this, few studies have been directly related to this topic because of the difficulty of recognizing clinical citations among biomedical papers (Li et al., 2022). Therefore, identifying clinical papers is a prerequisite for analyzing clinical citations of biomedical papers.

Biomedical papers whose article types are clinical trials or clinical guidelines are the most widely used source for counting clinical citations (Li et al., 2022). On the one hand, clinical guidelines are statements developed by domain experts to assist practitioners and patients with health-related decisions for specific diseases. Once a biomedical paper is cited in clinical guidelines, we may conclude that the paper has a direct impact on health practice (Yue et al., 2014). Compared with traditional bibliometric indicators, such as scientific citations and Mendeley statistics, clinical citation from clinical guidelines has been demonstrated to be a better representation of the clinical value of biomedical papers (Eriksson et al., 2019; Grant et al., 2000; Kryl et al., 2012; Thelwall & Kousha, 2016). On the other hand, being cited by clinical trials could also demonstrate the clinical impact of biomedical papers on clinical practice because the clinical trial is an essential and necessary channel through which biomedical discoveries can be finally applied to health practice (Kane et al., 2022; Thelwall & Kousha, 2016). Therefore, in previous studies, clinical citation has been defined as the citations received from clinical guidelines or clinical trials. For example, Hutchins et al. (2019) predicted the probability of biomedical papers being cited by clinical guidelines or trials for quantifying their translational progress using a random forest algorithm. Li et al. (2022) designed a multilayer perceptron neural network to predict long-term clinical citation counts for biomedical papers. In the current paper, we also adopted this method to count clinical citations of biomedical papers.

Besides the article type, scholars developed other approaches to identify clinical papers. For example, Narin et al. (1976) manually classified biomedical journals into four research levels: clinical observation, clinical mix, clinical investigation, and basic research. Then, papers published in clinical observation journals could be considered clinical papers. Lewison and Paraje (2004) made the journal classification into a semi-automatized process by using clue words in the titles and abstracts of biomedical papers. Weber (2013) initially proposed a MeSH-based triangle of biomedicine to classify biomedical papers into seven categories, in which human-related papers could be viewed as clinical papers. Hutchins, Davis, et al. (2019) and Ke (2019) modified Weber's triangle using fraction counting and word embeddings. Ke (2019) also proposed an indicator called "appliedness" (also named level score), which can be used for filtering clinical papers. With the level score, he further explored the citation inequity between clinical papers and basic papers (Ke, 2020). Li et al. (2023) designed a bibliometric measure called "Translational Progression" to quantitatively track the biomedical papers along the translational continuum by using biomedical knowledge representation at both entity and document levels.

Overall, most of the previous studies focused on how to identify clinical papers from biomedical papers or how to measure the "basicness" and "appliedness" of biomedical papers, which formed a strong foundation for the identification and analysis of clinical citations. However, few studies have yet explored the distribution or time characteristics of the clinical citation of biomedical papers. We still don't know how biomedical papers get clinically cited, or how they accumulate their clinical citations in their life cycles, which could be helpful for us to promote the clinical translation of biomedical research. Therefore, in the current study, we aimed to fill this research gap by retrospectively examining the temporal characteristics of biomedical papers accumulating clinical citations with a large-scale dataset.

## 3 Data and methods

### 3.1 Data collecting and pre-processing

The dataset used in this study was first downloaded in XML format from the official website of PubMed (https://pubmed.ncbi.nlm.nih.gov/) in October 2022. Then, we developed an XML parser with Java to extract bibliographic information for each paper, including PMID, article type, publication year, and MeSH terms. To get comprehensive citation relationships between articles, we integrated the PMID-PMID citation pairs from three open datasets, including the NIH Open Citation Collection (Hutchins, Baker, et al., 2019), the PubMed Knowledge Graph (Xu et al., 2020), and the Microsoft Academic Knowledge Graph (Färber and Ao, 2022; Wang et al., 2020). First, as the NIH Open Citation Collection has been integrated into the NIH iCite (Hutchins, davis, et al., 2019), we downloaded the whole iCite dataset in CSV format from its website (https://icite.od.nih.gov/) and adopted a Python parser to get the PMID-PMID citation pairs. These pairs included the citation relationships harvested from MedLine, PubMed Central, and CrossRef. Second, we downloaded the PubMed Knowledge Graph (version 4.0) from its website (http://er.tacc.utexas.edu/datasets/ped) and hosted it in a local MySQL database. Then, the PMID-PMID citation pairs of PubMed papers can be obtained from "C04_ReferenceList", which contained the NIH Open Citation Collection and the WOS citation data. Third, we also downloaded the MAKG datasets (https://makg.org/rdf-dumps/) and filtered the papers with PubMedID (i.e., PMID) and their references to obtain the PMID-PMID citation pairs. We integrated the three parts of PMID-PMID citation pairs and removed the duplicate data and some incorrect citation pairs. Finally, in our dataset, there were 34,017,389 biomedical papers published from 1940 to 2022, as shown in Fig.1A. Inspired by previous studies (Hutchins, Davis, et al., 2019; Li et al., 2022;), in this study, we adopted papers whose article type were clinical guidelines or clinical trials as the source of clinical citations. The distribution of clinical citations of these papers was illustrated in Fig.1B, which showed an apparent long-tail distribution. Over 80% of PubMed papers have no clinical citations, and about 90% of PubMed papers have less than two clinical citations.

As shown in Fig.1C, there were a total of 23,285,183 papers published in 1940-2013, and only 4,785,141 (20.55%) papers have been clinically cited. According to the number of clinical citation counts, we divided the 4,785,141papers into three groups; specifically, papers with at least 200 clinical citations were defined as highly clinically cited papers, while papers that had been clinically cited less than ten times were defined as lowly clinically cited papers. Those papers with clinical citations between 10 and 200 were defined as medium clinically cited papers. As illustrated in Fig. 1D and Table 1, 89.88% of papers with low, medium, and high clinical citations accounted for 89.884%, 10.063%, and 0.053%, respectively. Note that we did not consider papers with zero clinical citations in this study.

Based on the MeSH terms and article types, we also divided the 4,785,141 papers into four categories, including AC papers, H papers, Mixed papers, and Clinical guidelines or clinical trials. First, we classified all the 4,785,141 papers into three categories based on the triangle of biomedicine proposed by (Weber, 2013), in which the categories of biomedical papers were represented by the combination of three kinds of Medical Subject Headings (MeSH), i.e., the Animal-related (A), the Cell/molecular-related (C) and the Human-related (H) MeSH. Therefore, in this study, papers assigned with only H MeSH were classified as H papers, whose research content was more orientated to clinical science; papers assigned with only A, only C, or both A and C MeSH were AC papers, whose research content were closer to basic science; and papers assigned with AH, CH or ACH MeSH were Mixed papers. Second, clinical guidelines or clinical trials (CGCT) were identified solely based on their article types. CGCT papers had been widely used as the source of clinical citations of biomedical papers in previous studies (Hutchins, Davis, et al., 2019; Li et al., 2022; Li & Tang, 2021). The statistical information of these four categories of biomedical papers is illustrated in Table 1. Note that over 80% of CGCT (568,743) were also classified into H papers, this was because most clinical guidelines or clinical trials were human-related; however, over 94.9% of H papers were not CGCT. Although the overlapping, there will be no impact on the subsequent analysis.

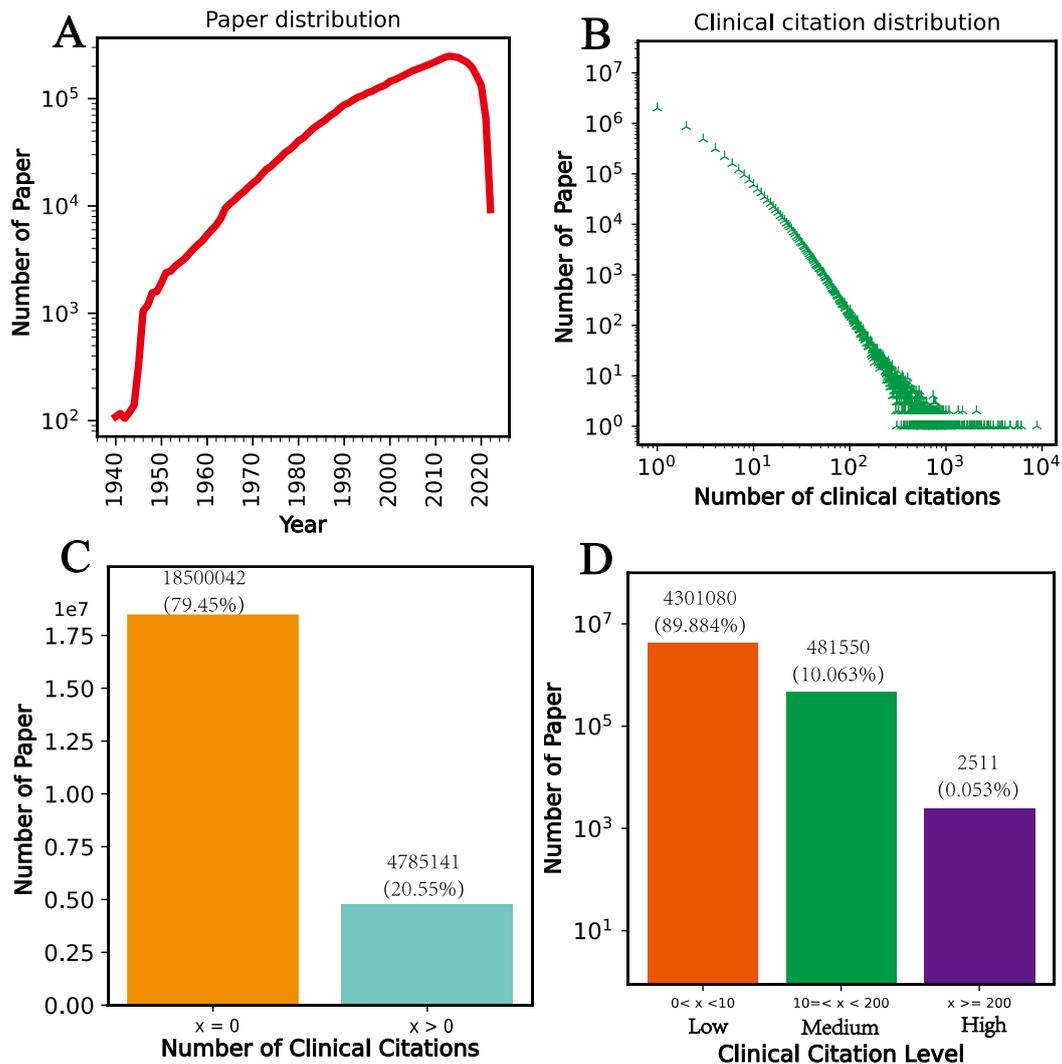

Fig.1 Descriptive information of biomedical papers in PubMed. (A) Changes in the number of biomedical papers in PubMed over years (1940-2022). (B) Distribution of clinical citations of biomedical papers (1940-2022). (C) Descriptive information on two groups of Biomedical papers (1940-2014). (D) Descriptive information on three groups of biomedical papers with at least one clinical citation (1940-2014). ["x": the number of clinical citations.]

We further analyzed the clinical citation length of biomedical papers. Specifically, we defined a paper's "clinical citation length" as the period from its publication year to the year it received its last clinical citation. For instance, the paper (PMID: 1202204) was published in 1975 and received its last clinical citations in 2022; its clinical citation length was 47 years. The average and maximum values of the clinical citation length of all PubMed papers over years were 5.64 and 8.93 years, as shown in Appendix 1. Thus, limiting the time frame of our analysis between 1940 and 2013 is appropriate to reduce the influence of the time cumulative nature of clinical citations.

Table1 Statistical information about biomedical papers in PubMed (1940-2013) from two perspectives: (1) clinical citation level and (2) translational science (i.e., ACH categories).

| Category / Clinical citation level | AC Papers | | | Mixed papers | | | H (Human) papers | Clinical Guidelines or Clinical Trials |
|---|---|---|---|---|---|---|---|---|
| | A (Animal) | C (Cell/ molecular) | AC | AH | CH | ACH | | |
| High | 9 | 7 | 9 | 40 | 148 | 36 | 2,191 | 930 |
| Medium | 9,537 | 1,419 | 9,824 | 18,152 | 64,796 | 14,083 | 360,371 | 141,479 |
| Low | 26,1740 | 44,341 | 317,000 | 165,859 | 614,825 | 209,765 | 2,624,738 | 376,702 |
| Zero | 1,643,638 | 1,280,099 | 2,417,564 | 570,265 | 1,647,617 | 781,395 | 8,203,797 | 179,444 |
| *Total* | *5,985,187* | | | *4,086,981* | | | *11,191,097* | *698,555* |

**3.2 Methods**

*3.2.1 Initial stage and build-up stage*

For a specific biomedical paper, we here divided its process of receiving clinical citations into two stages, including (1) the initial stage (from zero to one), which refers to a period from its publication year to the year when it receives its first clinical citation; and (2) the build-up stage (from one to N), which refers to a period after it receives its first clinical citation.

*3.2.2 Indicators for quantifying the process of receiving clinical citations*

To quantify the process of receiving clinical citations of biomedical papers, we used three indicators: (1) accumulative time, (2) interval time, and (3) response time. These three indicators were first proposed by Huang et al. (2019) to explore the length of time required for papers in computer science to receive their citations. Here, we used them for analyzing the clinical citations of biomedical papers. The details of the modified indicators are as follows:

**Accumulative Time (AT).** For a specific biomedical paper $P$, we denoted its publication year and its clinical citation-sequence as $PY_0$ and $[CC_1, CC_2, CC_3, ..., CC_i, ... CC_N]$, in which $CC_i$ was the $i^{th}$ clinical citations $P$ received after its publication. We denoted the publication year of $CC_i$ as $PY_i$ ($PY_0 \leq PY_1 \leq PY_i \leq PY_N, and\ 1 \leq i \leq N$). Then, the accumulative time ($AT_i$) can be defined as years required for $P$ to receive its $i^{th}$ clinical citations from its publication year, and can be calculated by:

$$AT_i = PY_i - PY_0 \quad (1)$$

**Interval Time (IT).** For a given biomedical paper $P$, the interval time referred to years required for receiving two adjacent clinical citations. Then, for the $(i-1)^{th}$ and $i^{th}$ clinical citations, the interval time between them, as denoted as $IT_i$, was given by:

$$IT_i = PY_i - PY_{i-1} \quad (2)$$

Noted that, if $CC_i$ and $CC_{i-1}$ were published in the same year (i.e., $PY_i = PY_{i-1}$), the interval time will be equal to 0. The value of interval time can be used for quantifying the difficulty in different stages of the process of clinical citations, including from publication to the first clinical citation, from the first to the second, …, and from $(i-1)^{th}$ to $i^{th}$. From this perspective, the value of interval time can also be used for assessing the translational difficulty of biomedical research.

**Response Time (RT).** We can easily find that when $i$ equals 1, $AT_1 = IT_1 = PY_1 - PY_0$, which represents the years it takes for a paper to receive its first clinical citations from publication, i.e., the initial stage. Here we defined $AT_1$ and $IT_1$ as the response time of the paper $P$. Response time is an important indicator because the status of the paper will be transformed from "unseen" to

"seen" (Egghe, 2000; Huang et al., 2019). Meanwhile, from the perspective of translational medicine, the knowledge in this paper will first flow into or be used by clinical science once it is clinically cited (Hutchins et al., 2019; Li et al., 2022).

*3.2.3 Empirical analysis*

With over 4.78M biomedical papers, we first conduct the empirical study from two perspectives: (1) the clinical citation level and (2) translational science (i.e., ACH categories). From the perspective of clinical citation level, we partitioned all papers into three groups based on the number of clinical citations, including highly, medium, and lowly clinically cited papers. In contrast, from the perspective of translational science, we divided all papers into four categories (i.e., AC papers, H papers, Mixed papers, and Clinical guidelines and clinical trials) according to their research topics and contents. For each perspective, we examined the probability distribution (PD) and cumulative probability distribution (PCD) of their response time (i.e., the years required for a paper to receive its first clinical citation from its publication) in its initial stage. Then, in the build-up stages, we investigated the PD and PCD of the years required for a paper to receive their third clinical citation and ninth clinical citation, i.e., from the 1st to the 3rd and from the 1st to the 9th, respectively.

Further, we divided the clinical citations of each biomedical paper into five zones, including 0%-20%, 21%-40%, 41%-60%, 61%-80%, and 81%-100%. For example, if a paper has received 100 clinical citations, its five zones are 0-20th, 21-40th, 41-60th, 61-80th, and 81-100th clinical citations, respectively. For each zone, we visualized and compared the mean, median, and mode of the average interval time (i.e., $IT_i$) of clinical citations for different groups of biomedical papers from the two perspectives.

## 4 Results

### 4.1 Initial stage: from 0 to 1st

*4.1.1 Clinical Citation Level Perspective*

Fig. 2 A illustrated the probability distribution of the response time of biomedical papers with different clinical citation levels, from which we found that, as the response time increased, all three curves exhibited a noticeable increase and then a sharp decline. The thresholds of the turn points for both highly and medium clinically cited papers were one year (in blue and red) and two years for the lowly clinically cited papers (in green). About 35% and 11% of the highly- and medium-clinically-cited papers received their first clinical citation within their publication year. In comparison, only about 3%, 10%, and 14% of lowly clinically cited papers received their first clinical citation in their publication year, one year, and two years after, respectively. Meanwhile, over three-quarters of the highly clinically cited papers received their first clinical citation within one year after publication. After the turn points, we also found that the curve of highly clinically cited papers dropped fastest, followed by the medium clinically cited papers, and the lowly clinically cited papers declined with the lowest speed.

The cumulative probability distribution of the response time of biomedical papers with different clinical citation levels is illustrated in Fig. 2B, in which all three curves displayed steep hillsides, indicating that most papers with clinical citations received their first clinical citations in a relatively short time. The curve of highly clinically cited papers had the fastest speed of increase, followed by that of the medium clinically cited papers, which was much faster than that of lowly clinically cited papers. Notably, only 37.2% of lowly clinically cited papers received their first clinical citation 3 years post-publication. In comparison, 91.2% and 82.9% of highly and medium clinically cited papers got their first clinical citation at the corresponding time. Nevertheless, the differences between the speeds of the three curves became smaller when the response time increased. For instance, when the response time increased to 20 years, 99.9%, 99.6%, and 93.8% of highly, medium, and lowly clinically cited papers received their first clinical citation.

We also analyzed the average years required for 80% of papers with different clinical citation levels to receive their first clinical citation, as shown in Fig. 2B (red dotted line) and Fig. 2C. The average length of time required for 80% of highly clinically cited papers to receive their first clinical citation was 1.380 years. In contrast, medium and lowly clinically cited papers were 2.761 and 10.285 years, respectively. This finding demonstrated again that the speed of receiving the first clinical citation satisfied the following order: highly clinically cited papers > medium clinically cited papers > lowly clinically cited papers. As papers with low and zero clinical citations accounted for about 97.92% of papers, this finding also proved the difficulty of clinical translation of biomedical research. When comparing these results with Huang et al. (2019), we found that a biomedical paper would take a much longer time to receive its first clinical citation than a CS paper to receive its first general citation.

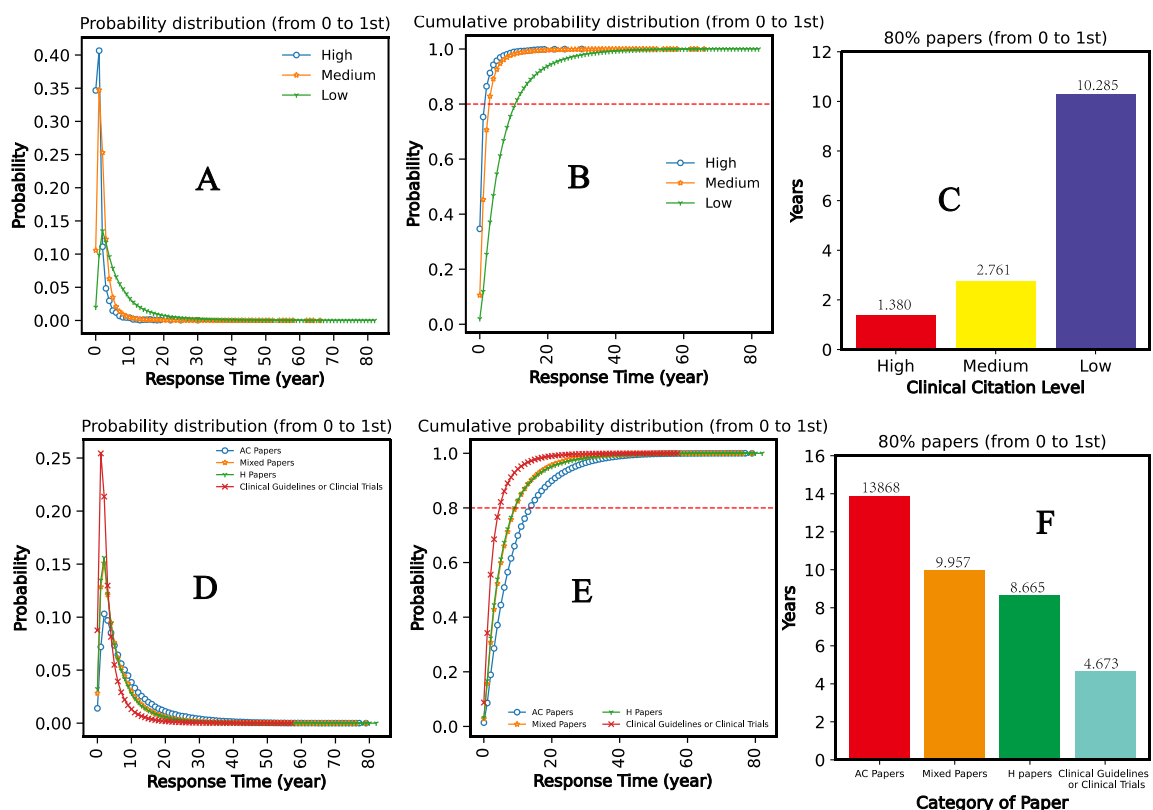

Fig.2 Probability distributions of response time for biomedical papers with different clinical citation levels (i.e., high, medium, and low) to receive their first clinical citation from publication (**A**, **B**). Comparisons of the time required for 80% of biomedical papers with different clinical citation levels to receive their first clinical citation from publication (**C**). Probability distributions of response time for four categories of biomedical papers (including AC papers, Mixed papers, H papers, and Clinical Guidelines or Clinical Trials) to receive their first clinical citation from publication (**D**, **E**). Comparisons of the time required for 80% of biomedical papers in different categories to receive their first clinical citation from publication (**F**).

*4.1.2 Translational Science Perspective*

In Fig. 2D, E, and F, biomedical papers were divided into four categories from the perspective of translational science (Hutchins, Davis, et al., 2019; Li et al., 2022; Li & Tang, 2021; Weber, 2013), including AC papers, Mixed papers, H papers, and Clinical guidelines and clinical trials (CGCT). Fig. 2D shows the probability distribution of response time for these four categories of biomedical papers, in which we can find that the four curves have similar trends as those in Fig. 2A, exhibiting an apparent increasing-decreasing trend. The turning point of CGCT was 1-year post publications, and those of the other three categories of papers were all two years post-publication. Particularly, ~9% of CGCTs have received their first clinical citations within their publication year, and more than 25% of GCGL 1-year post-publication. Although the percentages were smaller than those of highly clinically cited papers in Fig. 2A, they were much larger than those of the other three categories: ~2% and ~7% for AC papers, ~3% and ~13% for Mixed papers, and ~4% and ~14% for H papers, respectively. After the turn points, the decline speeds of the four curves also met the following order: CGCT > H papers > Mixed papers > AC papers.

Fig2. E illustrated the cumulative probability distribution of the response time of the four categories of biomedical papers. The four curves also showed steep slopes, indicating that most papers in all four categories received their first clinical citation relatively quickly. We observed that the curve for CGCT was on the far left, followed by the curves for H papers and Mixed papers, and the curve for AC papers was on the far right. This finding illustrated that the increasing speeds of the curves met the following order: CGCT > H papers > Mixed papers > AC papers. For example, 86.2 % of CGCTs received their first clinical citations when the response time was five years post-publication. At the same time, the percentages of H, Mixed, and AC papers were 67.5%, 66.2%, and 50.96% at the corresponding time, respectively. Similarly, the differences between different categories of papers were getting smaller when the response time increased. When the response time equaled 20 years post-publication, 99.01%, 95.13%, 95.72%, and 90.00% of CGCT, H papers, Mixed papers, and AC papers received their first clinical citation, respectively. Besides, we observed that the curves for H papers and Mixed papers almost overlapped, illustrating that the differences in response time between the two categories of papers were small.

Further, we investigated the average time required for 80% of these four categories of papers to receive their first clinical citations, as shown in Fig.2E (red dotted line) and Fig. 2F. The length of years required for 80% of AC papers to receive their first clinical citation was 13.868 years, while that of CGCT was only 4.673 years; besides, the years required for 80% H papers and mixed papers to get their first clinical citations were 8.655 and 9.957 years, respectively. These findings demonstrated that papers with research content closer to clinical science (Such as clinical guidelines, clinical trials, and H papers) could receive their first clinical citation more quickly than basic papers (i.e., AC papers).

Based on the above observations, we can have two interesting conclusions. First, at the clinical citation level perspective, highly clinically cited papers have a distinct advantage over medium and lowly clinically cited papers in the time required to receive their first clinical citations. Second, from the translational science perspective, papers closer to clinical science can receive their first clinical citations more quickly than those closer to basic science.

**4.2 Build-up stage: from 1st to Nth**

To examine the process of the build-up stage of clinical citations for biomedical papers, we designed 2 case studies that included clinical citations received from 1st to 3rd and from 1st to ninth. We chose nine as a research proxy because it was the threshold between lowly and medium clinically cited papers (see Fig. 1D). Besides, the analysis in this part was also conducted from the two perspectives, i.e., clinical citation level perspective and translational science perspective.

*4.2.1 Clinical Citation Level Perspective*
Fig. 3A and D showed the probability distribution of the time required for biomedical papers with different clinical citation levels to receive clinical citations from 1st to 3rd and from 1st to 9th. In Fig.3 A, it was observed that the curve for highly clinically cited papers declined sharply when the time was less than ten years and then remained zero, indicating that all highly clinically cited papers had received their third clinical citations within ten years after their first clinical citations. Particularly, ~50% and ~39% of highly clinically cited papers had received their third clinical citations in the same year and one year after their first clinical citations, respectively. Differently, the other five curves in Fig. 3A and D all exhibited increasing-decreasing trends, although the turn points occurred in various years. When comparing the identical groups in the two figures, we observed that highly clinically cited papers received their third and ninth clinical citations quicker than the other two groups.

Fig. 3B and E illustrated the cumulative probability distribution of the time required for biomedical papers with different clinical citation levels to receive clinical citations from 1st to 3rd and from 1st to 9th. It was observed that all six curves showed steep slopes; meanwhile, in both figures, the curve for highly clinically cited papers was on the far left, followed by the curve for the medium group, and the curve for the lowly group was on the far right. We also observed that the distances among curves in Fig. 3B were much smaller than those in Fig. 3E. For instance, within two years after receiving the first clinical citation, ~96%, ~77%, and ~30% of highly, medium, and lowly clinically cited papers were expected to receive their third clinical citations, respectively; the corresponding percentages were ~76%, ~10% and ~0% in the case of the ninth clinical citations. These findings again demonstrated that highly clinically cited papers received clinical citations much quicker than medium and lowly clinically cited papers in the build-up stage.

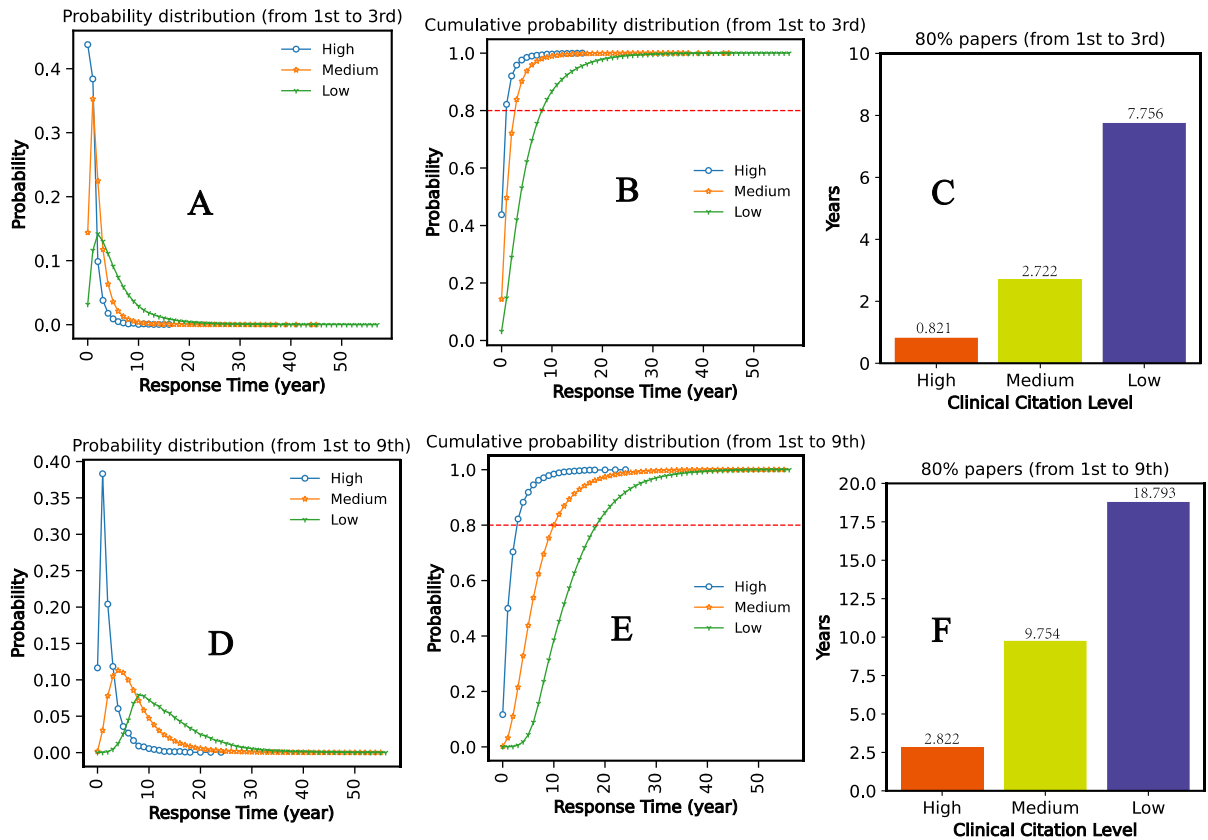

Fig.3 Probability distributions of response time for biomedical papers with different clinical citation levels (i.e., high, medium, and low) to receive clinical citations from 1st to 3rd (**A**, **B**). Comparisons of the time required for 80% of biomedical papers with different clinical citation levels to receive clinical citations from 1st to 3rd (**C**). Probability distributions of response time for biomedical papers with different clinical citation levels to receive clinical citations from 1st to 9th (**D**, **E**). Comparisons of the time required for 80% of biomedical papers with different clinical citation levels to receive clinical citations from 1st to 9th (**F**).

The average time required for 80% of biomedical papers with different clinical citation levels to receive clinical citations from 1st to 3rd and from 1st to 9th were compared with each other in Fig. 3C and F. For example, in Fig. 3C, highly clinically cited papers averagely took 0.821 years to acquire their third clinical citations, while lowly clinically cited took 9.4 times longer (i.e., 7.756 years) to finish this process. The corresponding years were 2.822 and 18.973 (6.659 times longer) for highly and lowly clinically cited papers for acquiring clinical citations from 1st to 9th, respectively.

### 4.2.2 Translational Science Perspective

Fig. 4 A and D were the probability distributions of the required time for four categories of biomedical papers to receive clinical citations from 1st to 3rd and from 1st to 9th, in which all six curves exhibited an increasing-decreasing trend. However, the turn points occurred in different years. A similar pattern can be easily found when comparing the same category of papers in the two figures: the curves for H and Mixed papers were almost overlapped. In contrast, the curves for AC papers and clinical guidelines or clinical trials (CGCT) differed. Notably, one year after the first clinical citation, ~ 27% and ~ 5.7% of highly clinically cited papers received their 3rd and 9th clinical citations. In contrast, the corresponding percentages for lowly clinically cited papers were only ~12.5% and ~0.9%, respectively.

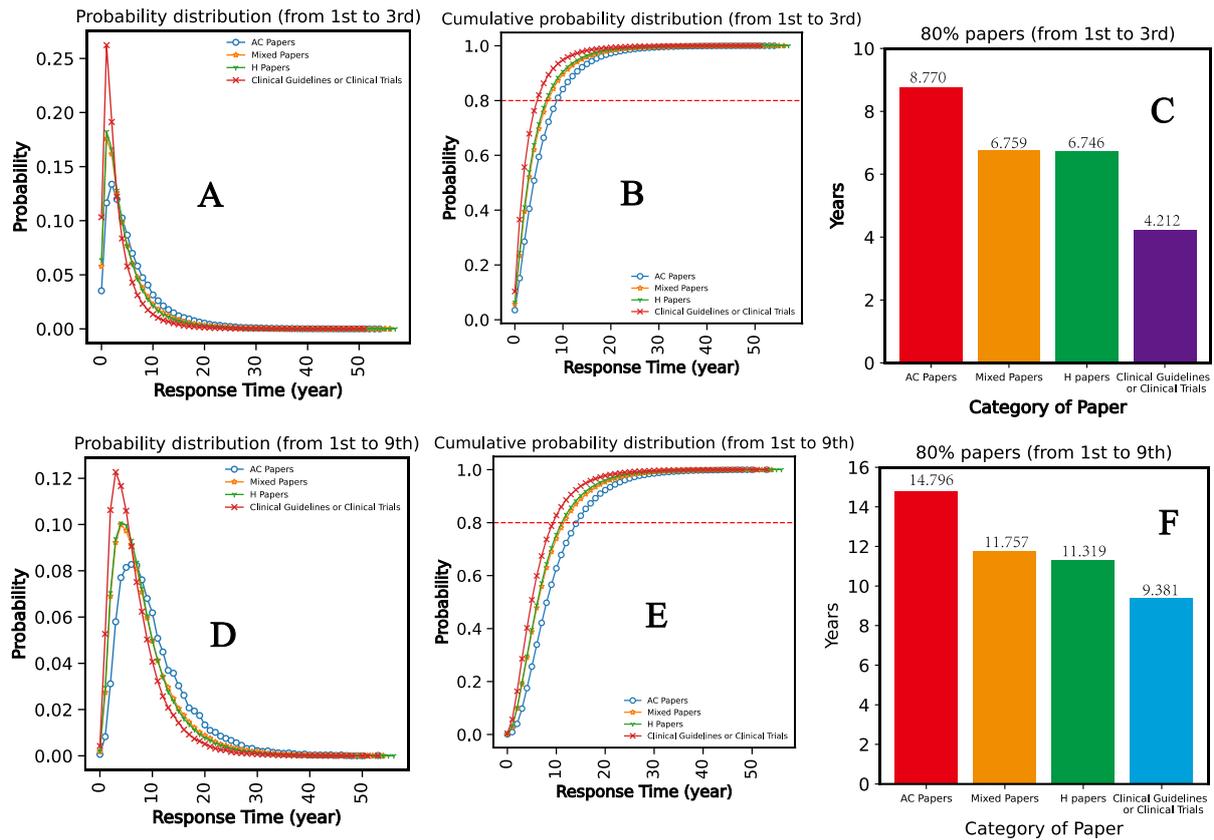

Fig.4 Probability distributions of response time for four categories of biomedical papers to receive clinical citations from 1st to 3rd (**A**, **B**). Comparisons of the time required for 80% of biomedical papers in four categories to receive clinical citations from 1st to 3rd (**C**). Probability distributions of response time for four categories of biomedical papers to receive clinical citations from 1st to 9th (**D**, **E**). Comparisons of the time required for 80% of biomedical papers in four categories to receive clinical citations from 1st to 9th (**F**).

The cumulative probability distributions of the required time for the four categories of biomedical papers from 1st to 3rd in Fig. 4B and from 1st to 9th in Fig. 4 showed a sharp increase. The increase speeds of the four curves in both figures satisfied the following order: Clinical guidelines or clinical trials (CGCT) > H papers > Mixed papers > AC papers. Notably, the curves for H papers and Mixed papers in both figures followed almost the same trajectory, although H papers had a slight speed advantage. The difference between CGCT and Ac papers in the two figures was the largest. For example, in the fourth year after the first clinical citation, ~78% of CGCT and ~50% of AC papers received their 3rd clinical citations (see Fig. 3B). The corresponding percentages were ~41% and ~27% for CGCT and AC papers to received their 9th clinical citations (see Fig. 3E).

Fig. 4C and F illustrated the average time required for 80% of the four categories of biomedical papers to receive their clinical citations from the 1st to the 3rd and from the 1st to the 9th. AC papers took the longest time in both figures, i.e., 8.770 years for receiving clinical citations from the 1st to the 3rd and 14.796 years from the 1st to the 9th, respectively; the corresponding time for CGCT were respectively 4.212 years and 9.381 years. Besides, the time required for Mixed papers and papers were very close to each other in both figures.

**4.3 Accumulation of clinical citations in five zones**

In the section above, we used the third and ninth clinical citations as two proxies to explore how biomedical papers in different groups accumulated their clinical citations over the years. In this section, we further examine the accumulation of clinical citations for biomedical papers in the five zones we divided.

*4.3.1 Clinical Citation Level Perspective*

Fig 5A, B, and C were plots of the interval time for biomedical papers with different clinical citation levels receiving clinical citations in five zones, in which the short horizontal lines in the top and bottom were respectively maximum and minimum values, the vertical lines represented the ranges of the average time of biomedical papers to receive the corresponding number of clinical citations, and the yellow, green and red curves were the changes in the mean, median and mode of the interval time over zones, respectively. It was observed that the lowly clinically cited papers had the largest span of interval times (blue vertical lines), followed by the medium paper group, and highly clinically cited papers had the most minuscule span. Meanwhile, no noticeable change patterns were observed over zones in all three figures for the interval period.

We compared the mean, median, and mode of the interval time of biomedical papers with different clinical citation levels in Fig 5D, E, and F. The blue line representing the mean interval time of lowly clinically cited papers was higher than the orange line in all five zones, which was overall higher than the green line (Fig 5D). This illustrates that the highly clinically cited papers received clinical citations much easier than medium clinically cited papers, which received clinical citations much easier than lowly clinically cited papers. Meanwhile, the median curves of interval time of the three groups in Fig. 5E also exhibited a similar pattern, i.e., the highly clinically cited papers received their clinical citations more quickly than the medium group of papers, and the lowly clinically cited papers had the most significant difficulty in receiving clinical citations. Finally, the three mode curves in Fig. 5F were slightly different, although the blue line representing the mode of interval time of lowly clinically cited papers was also the highest. Specifically, all the mode values of highly and medium clinically cited papers' interval time were nearly zero. However, the interval time of lowly clinically cited papers was about 2 years, 1 year, 1 year, 1 year, and 1 year in the five zones, respectively. This illustrated that even most lowly clinically cited papers acquired clinical citations relatively quickly.

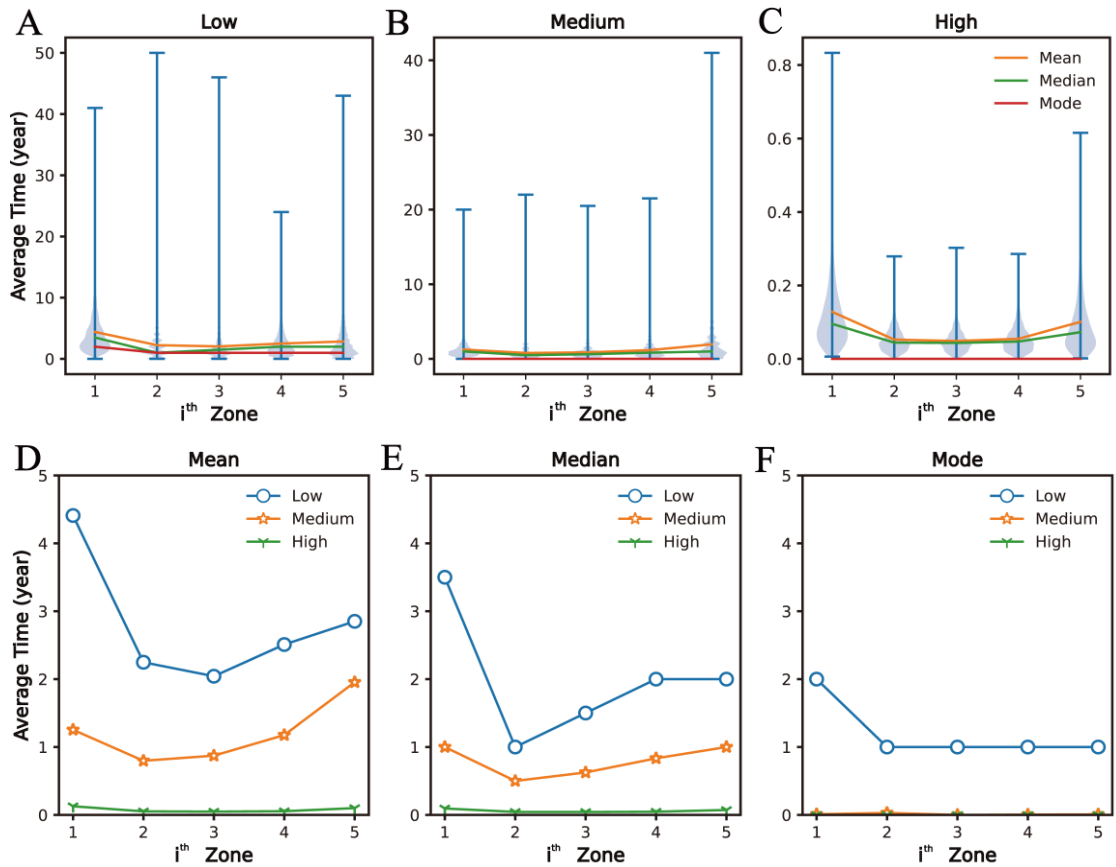

Fig.5 Plots of the average interval time for biomedical papers with different clinical citation levels receiving clinical citations in five zones (A, B, and C). The comparison of the mean (D), median (E), and mode (F) of the average interval time in each zone for biomedical papers with different clinical citation levels.

### 4.3.1 Translational Science Perspective

Fig 6A, B, C, and D illustrated the average interval time for four categories of biomedical papers in five zones. It was observed that the spans of the average interval time of all four categories of papers were very similar. Notably, in all four figures, the average interval time in the second zone (i.e., 21%-40%) was the longest, while that in the fourth zone (i.e., 61%-80%) was the shortest. This illustrated that the stage at which biomedical papers were most likely to receive clinical citations was from 21% to 40%, while the most challenging stage to receive clinical citations was from 61% to 80%. Meanwhile, for AC papers, the interval period of the first zone was more extended than that of the third zone, which was the opposite of the other three categories of biomedical papers. This again indicated that AC papers received their clinical citations at the beginning stage harder than other categories of papers. This phenomenon may be caused by the low visibility of AC papers among clinical scientists. The average interval time of papers in the first zone also met the following order: CGCT < Mixed papers < H papers < AC papers, consistent with the finding that CGCT can receive their clinical citation more quickly in the initial stage. However, when comparing Fig 6D with Fig 5C, we found that highly clinically cited papers have the absolute advantage of receiving clinical citations over all four categories of papers.

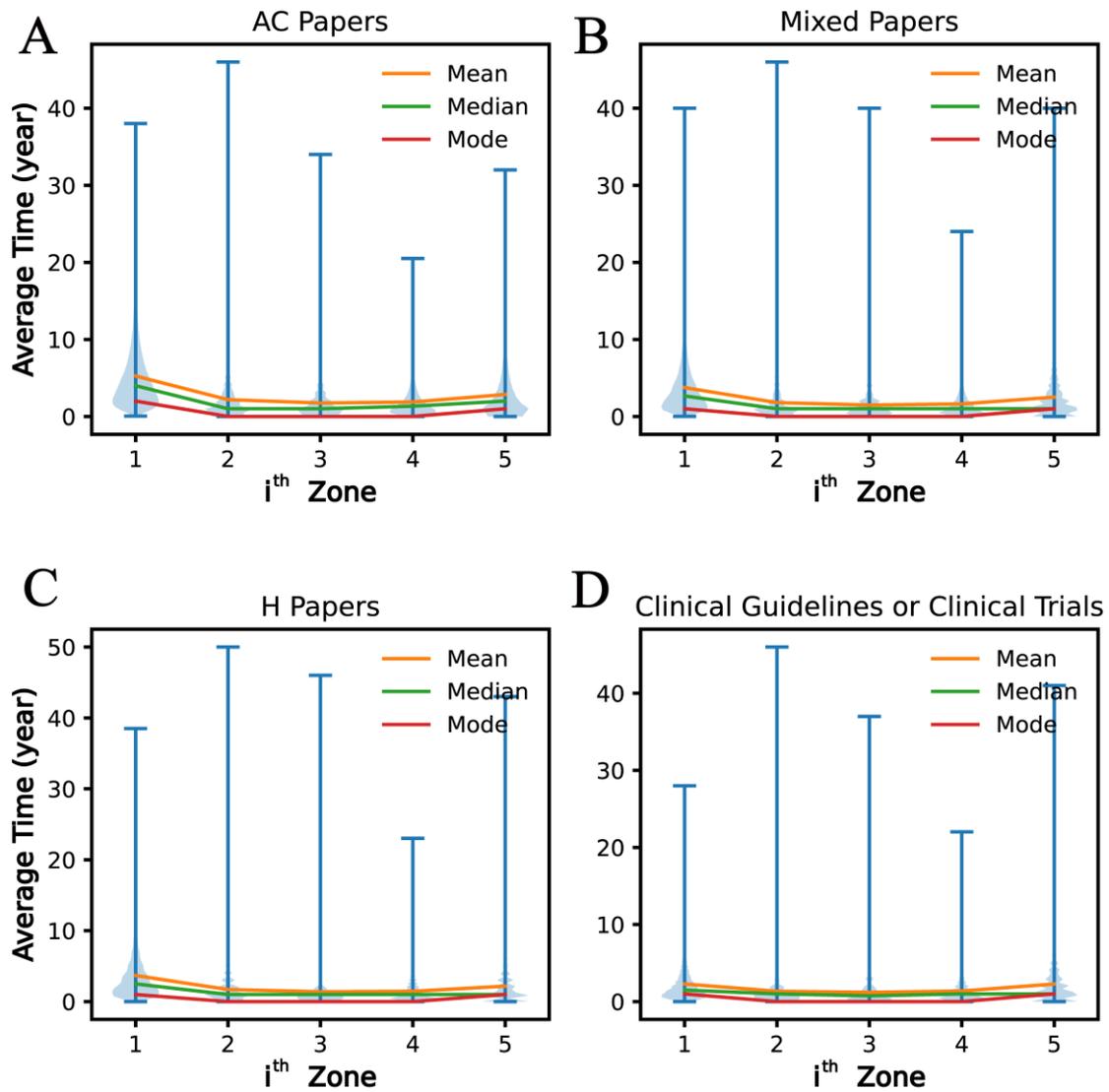

Fig.6 Plots of the average interval time for four categories of biomedical papers receiving clinical citations in five zones.

In Fig. 7, we compared the mean, median, and mode of the average interval time of four categories of biomedical papers in the five zones. It was observed that the blue curve representing the mean of the interval time of AC papers in five zones was highest, followed by the curves for Mixed papers and H papers, and the curve for CGCT was lowest. This demonstrated that CGCT received clinical citations more quickly than other categories of biomedical papers. The four median curves in Fig 7B also illustrate that AC papers took more years to receive clinical citations than other categories of biomedical papers, especially for the CGCT, because the median curve for AC papers was higher than the other three. Besides, the four mode curves displayed in Fig 7C were also illustrative. Specifically, in the first zone (i.e., 0-20%), the mode curve for AC papers was higher than those of other categories of biomedical papers, whose mode curves overlapped. This indicated that most AC papers received clinical citations harder than other categories of biomedical papers at the initial stage. Besides, after the first zone, all four curves overlapped and were less than 2 years in Fig 7C, indicating that most biomedical papers received their clinical citations relatively quickly.

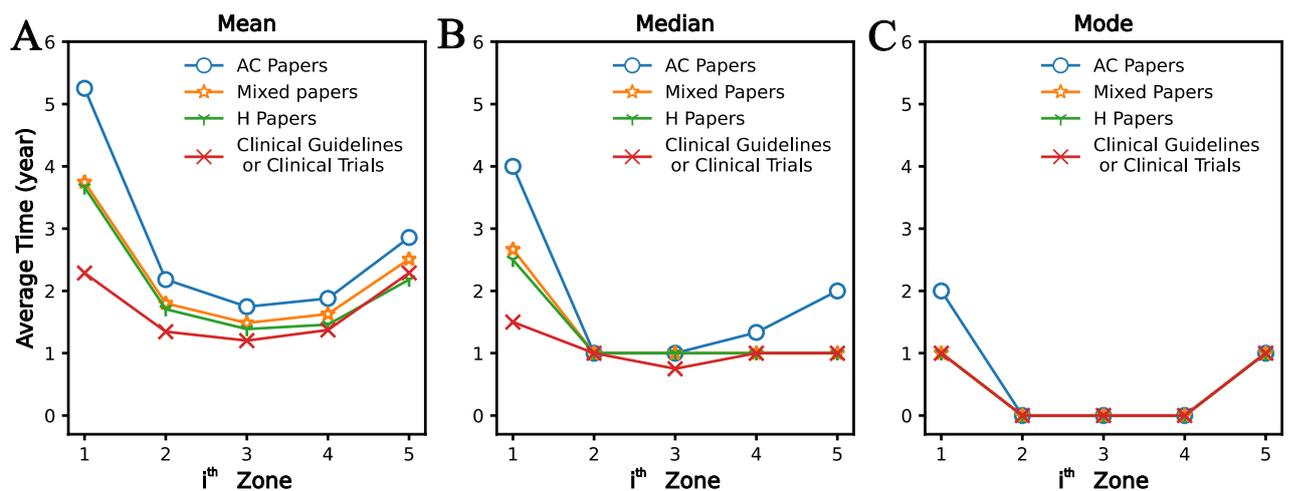

Fig.7 The comparison of the mean (A), median (B), and mode (C) of the average interval time in each zone for four categories of biomedical papers.

**5 Robustness analysis**

To enhance the reliability of the results in the current study, we conduct a robustness analysis from two aspects. First, we conduct a comparison analysis of time characteristics for biomedical papers between clinical citations and general citations. Second, we conduct an extension analysis on that, to what extent the citation count of basic papers needs to accumulate before they start receiving clinical citations, and whether this type of citation continues to accumulate over time.

**5.1 Comparison analysis between clinical citations and general citations**

When comparing with the results of Huang et al. (2019), we found that it was a greater challenge for biomedical papers receiving clinical citations than CS papers receiving citations. This finding raised a question that how biomedical papers have accumulated their citations from general literature. Were biomedical papers receiving citations from general literature easier than receiving clinical citations?

To answer the above questions, we defined general citations as the citations of biomedical papers receiving from general literature. We then analyzed and visualized the probability distribution (PD) and the probability cumulative distribution (PCD) of response time for biomedical papers to receive their citations from general literature, including two stages, i.e., from 0 to $1^{st}$ and from $1^{st}$ to $N^{th}$. Fig. S2, S3, and S4 in the Appendix Information displayed the results, which revealed several interesting findings. In the initial stage, the differences in the PD and PCD between all groups of biomedical papers receiving general citations (Fig. S2 ABDE) were much smaller than those receiving clinical citations (Fig. 2 ABDE). Highly clinically cited papers had the highest speed to accumulate their general citations, followed by the medium group, and the lowly group was slowest. However, in the perspective of translational science, the original order in Fig. 2 was reversed. Papers closer to clinical science had the slowest speed to accumulate their general citation, i.e., the speed of receiving first general citations: CGCT < H papers < Mixed papers < AC papers. Meanwhile, when comparing Fig. S2 CF with Fig. 2 CF, we found that it is much easier for biomedical papers to accumulate general citations than clinical citations. Biomedical papers at most took up 2.848 years to receive their first general citations (CGCT in Figure S2 F). This indicated that biomedical papers were easier to receive their first general citations than CS papers (Huang et al., 2019).

In the build-up stages (Fig. S3 and S4), all the curves for the PD and PCD of response time for biomedical papers to receive general citations were similar to those in Fig. 3 and 4, however, the differences between curves were much smaller. We can also observe that general citations were much easier for biomedical papers to receive than clinical citations in the build-up stages, including from $1^{st}$ to $3^{rd}$ and from $1^{st}$ to $9^{th}$. From the clinical citation level perspective, the speed of accumulating general citations from $1^{st}$ to $N^{th}$: highly clinically cited papers > medium clinically cited papers > lowly clinically cited papers, the order of which was consistent with that in clinical citations. In contrast, the order in perspective in translational science was reversed: CGCT < H papers < AC papers < Mixed papers. It can also be found that biomedical papers were easier to accumulate their general citations in the build-up stages than CS papers.

Overall, it was much easier for biomedical papers to accumulate citations from general literature (i.e., general citations) than from clinical guidelines or clinical trials (i.e., clinical citations) in both the initial and build-up stages. Biomedical papers with higher clinical citations could be more likely to have a higher speed of receiving general citations. Different from clinical citation, biomedical papers closer to clinical science had no advantages of receiving general citations. On the contrary, AC papers accumulated their general citations at the fastest speed, while CGCT was the slowest. Nevertheless, the differences in response time for accumulating general citations among different categories/groups of biomedical papers were relatively small compared to clinical citations. Besides, in both the initial and build-up stages, biomedical papers had an advantage of receiving general citations than CS papers compared to the corresponding results in Huang et al. (2019).

**5.2 Extension analysis on basic papers**

To better understand the clinical impact of basic papers, we conduct an extension analysis on basic papers. We defined basic papers as biomedical papers that were assigned with only Animal-related, only Cell/molecular-related, or both Animal-related and Cell/molecular-related MeSH terms, i.e., AC papers (in section 3.1). In our dataset, 25.7% (5,985,141) of biomedical papers were basic papers, and 645,166 (only 10.78%) of these basic papers have one or more clinical citations. Meanwhile, the first citations of the vast (99.15%) of these basic papers were not clinical. That is, most basic papers need to accumulate other types of citations before they start receiving clinical citations. Therefore, we further asked, to what extent the citation count of basic papers needs to accumulate before they start receiving clinical citations, and whether this type of citation continues to accumulate over time.

Fig. 8A and B showed the number distribution and cumulative probability distribution of the rank of the first citations in citation sequences of biomedical papers. We found that the number of papers decreased as the rank of the first citation increased. The second place (yellow bubble in the inset of Fig. 8A) had the greatest number of papers (over 40,000), accounting for about 7.29% (the first point in Fig. 8B) of those basic papers whose first citations were not clinical. Less than 40% basic papers had received their first clinical citation before they received 100 other types of citations (Fig. 8B). In terms of time, we found that the average rank of the first clinical citations of biomedical papers showed a decreasing trend over years, but there were significant fluctuations before the year 2000 (Fig. 8C). After 2000, the speed of this decreasing trend in the average rank was faster (the inset in Fig. 8C). Despite this, even in 2013, biomedical papers still needed to accumulated an average of more than 25 other types of citation before being clinically cited for the first time.

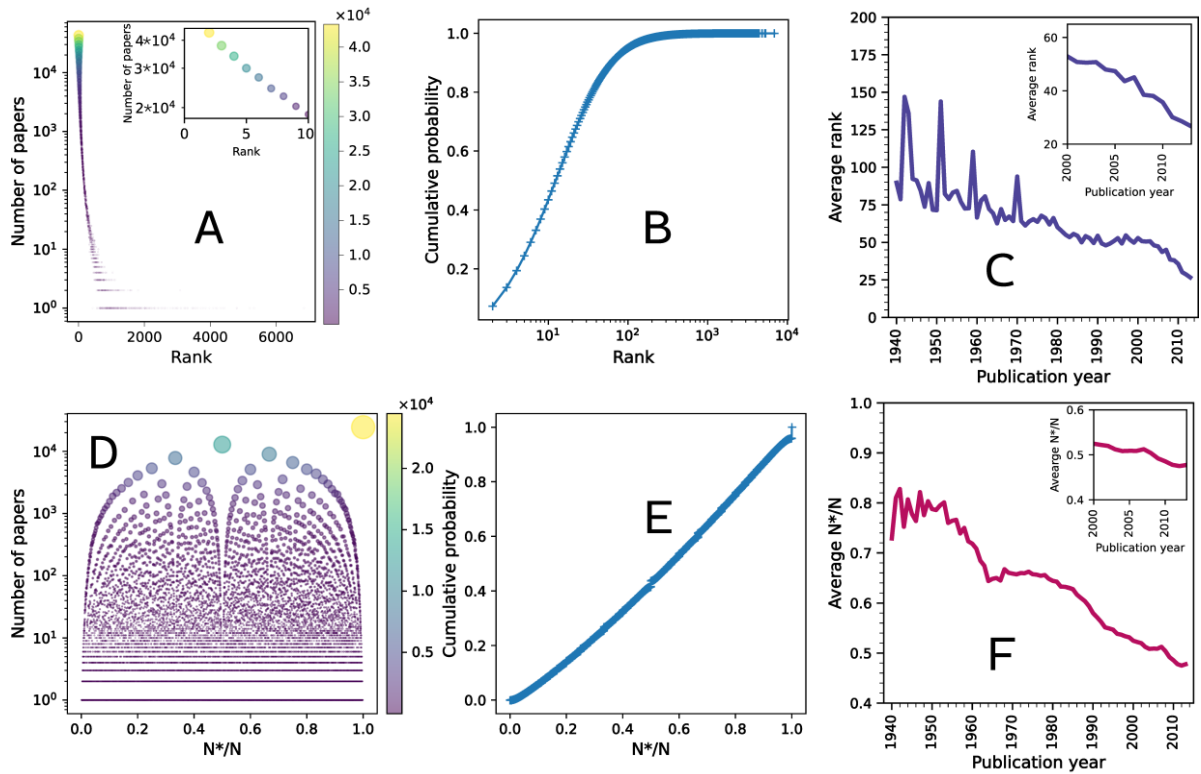

Fig. 8 The position of the first clinical citation in the citation sequences of basic papers. (AB) Number and cumulative probability distribution of the rank of the first citations in citation sequences of basic papers. (C) Changes in the average rank of the first clinical citations in citation sequences of basic papers over years. (DE) Number and cumulative probability distribution of the N*/N of basic papers. (F) Changes in the average of N*/N of basic papers over years.

The above findings inspired us to examine the relative position of the first clinical citations in citation sequences of basic papers by using N*/N, in which N* denoted the rank of the first clinical citation, and N represented the total number of citations. The value interval of N*/N was (0,1]. The larger the value of N*/N, the later the relative position of the first clinical citation in the citation sequence of a basic paper. Fig. 8 C and D showed the number distribution and the cumulative distribution of N*/N of basic papers, which revealed several interesting findings. First, the number of basic papers whose N*/N equaled to 1 was the largest (24470, 4.17%, yellow bubble in the upper right corner), indicating that their first clinical citations occurred at the end of the citation sequences (Fig. 8D). Meanwhile, according to the size and color of bubbles in Fig. 8D, it can be found that the values of N*/N of most papers were between 0.4 and 0.6. Fig. 8E showed that the cumulative probability distribution approximated a diagonal curve, indicating that about 50% of basic papers received their first clinical citations in the latter half of their citation sequences. As shown in Fig. 8F, the average values of N*/N also showed a decreasing trend with fluctuations over years, however, the N*/N was still over 0.47 in the year 2013. That is, basic papers averagely needed to accumulate approximately 47% citation counts before receiving their first clinical citations.

Further, we compared the proportion of different categories of citations for biomedical papers before and after receiving their first clinical citation, as shown in Fig. 9. Specifically, according to the categories of the citing papers, we classified citations into four categories: clinical citation, basic citation, mixed citation, and human-related citation (excluded the clinical citation). In Fig. 9A, the mean and median of the proportion of basic citations for basic papers were the largest (0.5658 and 0.5882, respectively), indicating that before being clinically cited, basic papers were more likely to be cited by basic papers. Meanwhile, the mode of the proportion of basic citations was 1, i.e., many basic papers (11.7%, N=69,561) were only cited by basic papers before receiving their first clinical citations. The mean and median of the proportion of mixed citations for basic papers were in second place, i.e., 0.2776 and 0.250.; and that of human-related citations for basic papers were the smallest. Therefore, we could conclude that basic papers were more likely to be cited by papers closer to basic science before they received their first clinical citations.

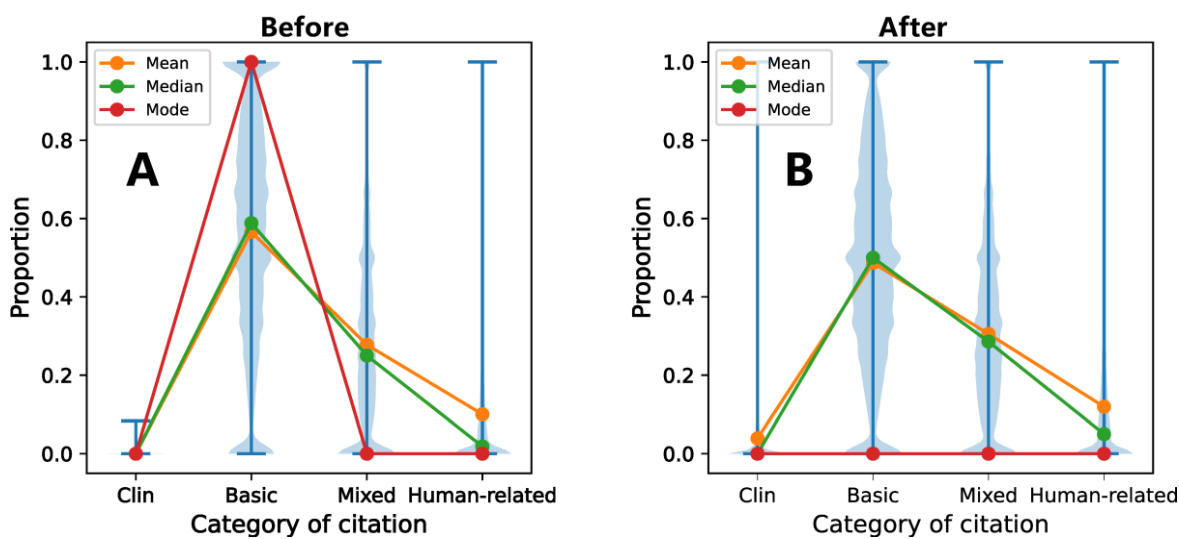

Fig.9 Plots of the proportion of the four categories of citations for basic papers before (A) and after (B) receiving the first clinical citations.

Fig. 9 showed the distribution of the proportion of different categories of citations for basic papers after they received their first clinical citations. We found that the mean and median of the proportion of basic citations were still the largest, followed by those of mixed and human-related citations, and those of clinical citations ranked the last place, that is, basic citations > mixed citations > human-related citations > clinical citations. This indicated that basic papers were still more likely to be cited by papers that were closer to basic science after they being clinically cited for the first time.

**6 Discussion and conclusion**

This study examined the time required for biomedical papers to receive their clinical citations from two perspectives, i.e., the clinical citation level (i.e., highly, medium, and lowly) and the translational science (i.e., ACH categories). 23,285,183 PubMed papers published between 1940-2013 were used as the dataset. From the clinical citation level perspective, biomedical papers were partitioned into three groups based on the number of their clinical citations, including highly, medium, and lowly clinically cited papers; from the translational science perspective, we partitioned biomedical papers into AC papers, Mixed papers, H papers, and Clinical guidelines or clinical trials, according to their article types and MeSH terms assigned. Three indicators (i.e., the Accumulative Time, the Interval Time, and the Response Time) were developed to quantify the time required for biomedical papers in the process of receiving clinical citations, which was divided into two stages, including the initial stage (i.e., from 0 to the 1st clinical citation) and the build-up stage (i.e., from the 1st to the Nth clinical citation). We also divided the clinical citations of a biomedical paper into five zones and then visualized and compared the mean, median, and mode of the average interval time of receiving clinical citations in each zone for different groups or categories of biomedical papers. To enhance the reliability of the results, we further conducted robustness analysis from two aspects, including (1) comparison analysis between clinical citations and general citations and (2) extension analysis on basic papers.

We had several interesting findings. First, from the perspective of clinical citation level, we found that highly clinically cited papers received their clinical citations much quicker than medium and lowly clinically cited papers, whether in the initial stage or the build-up stage. This result differed from the research of Huang et al. (2019) in the field of computer science, in which there was no noticeable difference among highly, medium, and lowly cited CS papers in the initial stage. Meanwhile, biomedical papers were easier to receive general citations than CS papers (Fig. S2, S3, and S4), however, the time required for biomedical papers to receive clinical citations was much longer than that for CS papers to receive the corresponding number of citations in both the initial and build-up stages. These findings indicated the great challenge for biomedical papers to receive clinical citations and the difficulties in the clinical translation of biomedical research. In the build-up stage, when the number of clinical citations increased, the length of time required for accumulating the corresponding number of clinical citations increased by a large margin; moreover, the difference in the time required among the three groups of biomedical papers also largely increased.

Second, from the perspective of translational science, it was found that the speed of receiving clinical citations for biomedical papers satisfied the following order in both the initial stage and build-up stage: AC papers < Mixed papers < H papers < Clinical guidelines or clinical trials, which was consistent with the progression of each category of biomedical papers on the translational continuum. This demonstrated that biomedical papers closer to clinical science (such as clinical guidelines or H papers) had the absolute advantage of receiving clinical citations over biomedical papers closer to basic papers (such as AC papers). This phenomenon may be because papers closer to clinical science were easier to be seen and cited by clinical scientists than basic papers; furthermore, the Matthew effect (i.e., accumulative advantage) could also make this advantage larger over time (Price, 1976). When comparing the results of the two perspectives, we found that highly clinically cited papers had the desperate advantage of receiving clinical citations over even the clinical guidelines or clinical trials.

Third, in robustness analysis, we found papers closer to basic papers were easier to receive general citations, which was the exact opposite of receiving clinical citations. This phenomenon could be explained by the citation disadvantage of clinical research, which was first discussed by Ke (2020). In the extension analysis, the results showed that, before being clinically cited for the first time, basic paper averagely needed to accumulate approximately 47% citation counts or 25 citation counts. These citation relationships built up a bridge between basic science and clinical science, which could be significant for analyzing the patterns in the clinical translation of biomedical research. It can also be found that basic papers were more likely to be cited by papers that were closer to basic science whether before or after they were clinically cited for the first time.

We here summarized the implications of the current study. First, our study revealed the temporal characteristics of how different categories of biomedical papers accumulate their clinical citations based on a large-scale dataset. The findings shed light on the time mechanisms of clinical translation of biomedical research, especially for basic research, which could be useful for research managers to have a better understanding of the clinical translation of biomedical research, and to make personalized and fine-grained scientific policies for different categories of biomedical research. Methodologically, the modified indicators in this study including the Accumulative Time, the Interval Time, and the Response Time, originally proposed by Huang et al., (2019), could be helpful for pharmaceutical companies and government policy-makers. They can directly use them to monitor and quantify the clinical translational lags of biomedical research in real-time. Meanwhile, the Response Time could be used as an entitymetric indicator for discovering biomedical research with the high potential to be clinically translated (Li et al., 2019; Li et al., 2023). What's more, the current research results could be interesting for young scholars in biomedicine to get more attention from clinical science and to obtain success in their scientific careers, especially for those in basic science.

The limitations of this study were as follows. First, we only considered the clinical guidelines and clinical trials as the source of clinical citations. However, biomedical papers might be cited by other clinical publications, such as clinical cases and other health policies. Second, in this study, we focused on the number of clinical citations but ignored the semantic information of clinical citations, such as their motivations, contexts, and citation functions. In the future, more related data sources could be used for counting the clinical citations of biomedical papers, such as health-related policy texts and patents. Meanwhile, NLP techniques could be applied to the full texts of biomedical papers, to understand the motivation, contexts, and citation functions of clinical citations. We also planned to explore the self-clinical citation behavior in future work based on the author name disambiguation algorithms of PubMed authors. Besides, a related topic we will explore in the future is to examine the causal factors affecting the accumulation of clinical citations of biomedical papers.

**Acknowledgments**

This work was supported by the National Natural Science Foundation of China (grant no. 72204090). This work was also supported by the Ministry of Education of Humanities and Social Science Project (grant no. 22YJC870014). The computation is completed in the HPC platform of Huazhong University of Science and Technology. We also thank the two anonymous reviewers for improving the quality of this article.

**Declarations**

**Conflict of Interest:** The authors have no relevant financial or non-financial interests to disclose.

to Improve Clinical Practice. Journal of the American Geriatrics Society, 62(4), 754–761.

**Appendix 1 The average of clinical citation ages of biomedical papers over years**

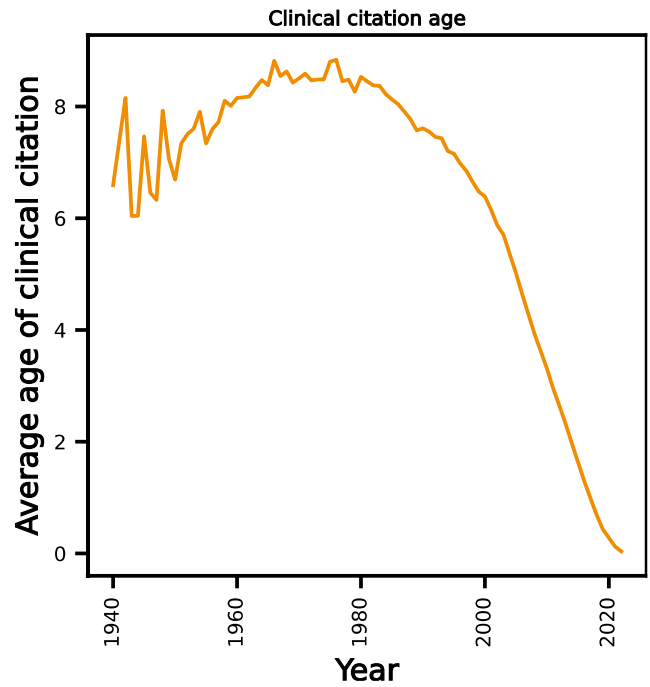

Fig. S1 The average of clinical citation ages of biomedical papers over years

**Appendix 2 The PD and PCD of response time for biomedical papers to receive their first citation from general literature in the initial stage (i.e., from 0 to 1st).**

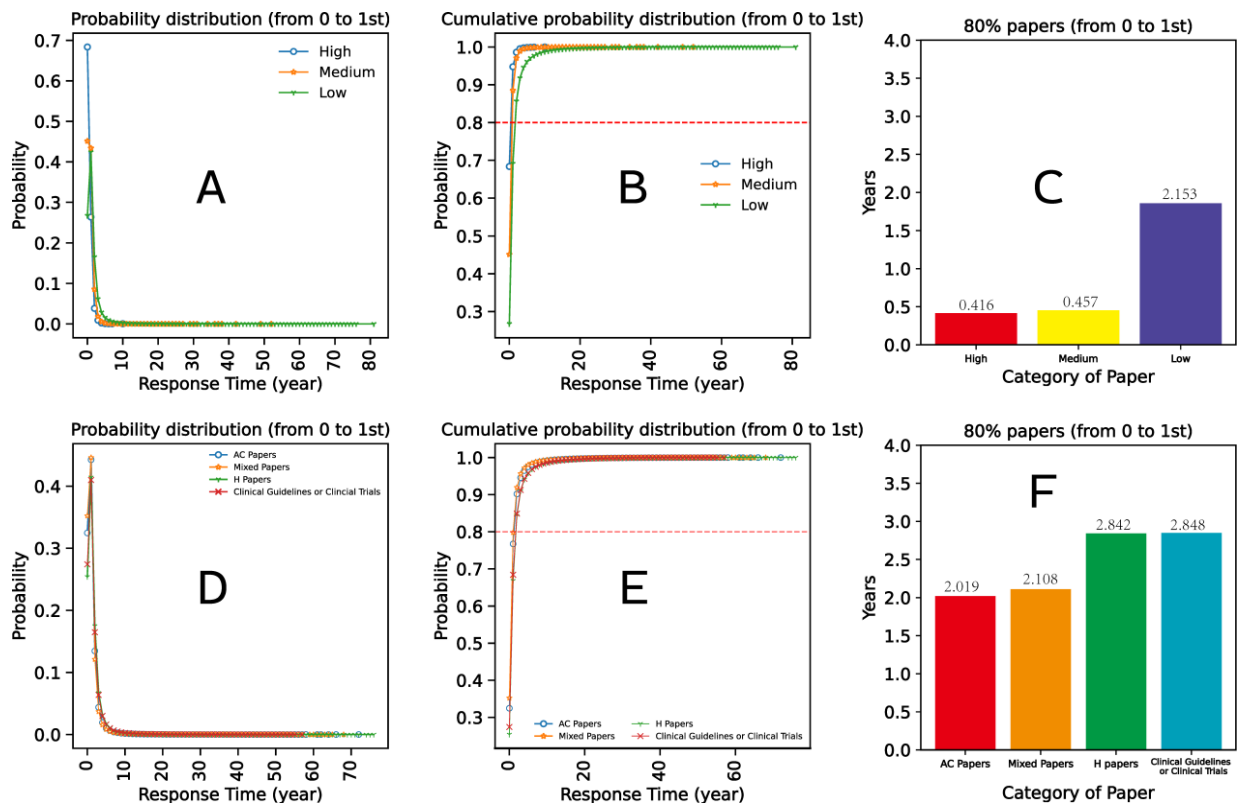

Fig.S2 Probability distributions of response time for biomedical papers with different clinical citation levels (i.e., high, medium, and low) to receive their first citation from general literature (A, B). Comparisons of the time required for 80% of biomedical papers with different clinical citation levels to receive the first citation from general literature (C). Probability distributions of response time for four categories of biomedical papers (including AC papers, Mixed papers, H papers, and Clinical Guidelines or Clinical Trials) to receive the first citation from general literature (D, E). Comparisons of the time required for 80% of biomedical papers in different categories to receive their first citation from general literature (F).

**Appendix 3 The PD and PCD of response time for biomedical papers with different clinical citation levels to receive their citation from general literature (from 1st to Nth).**

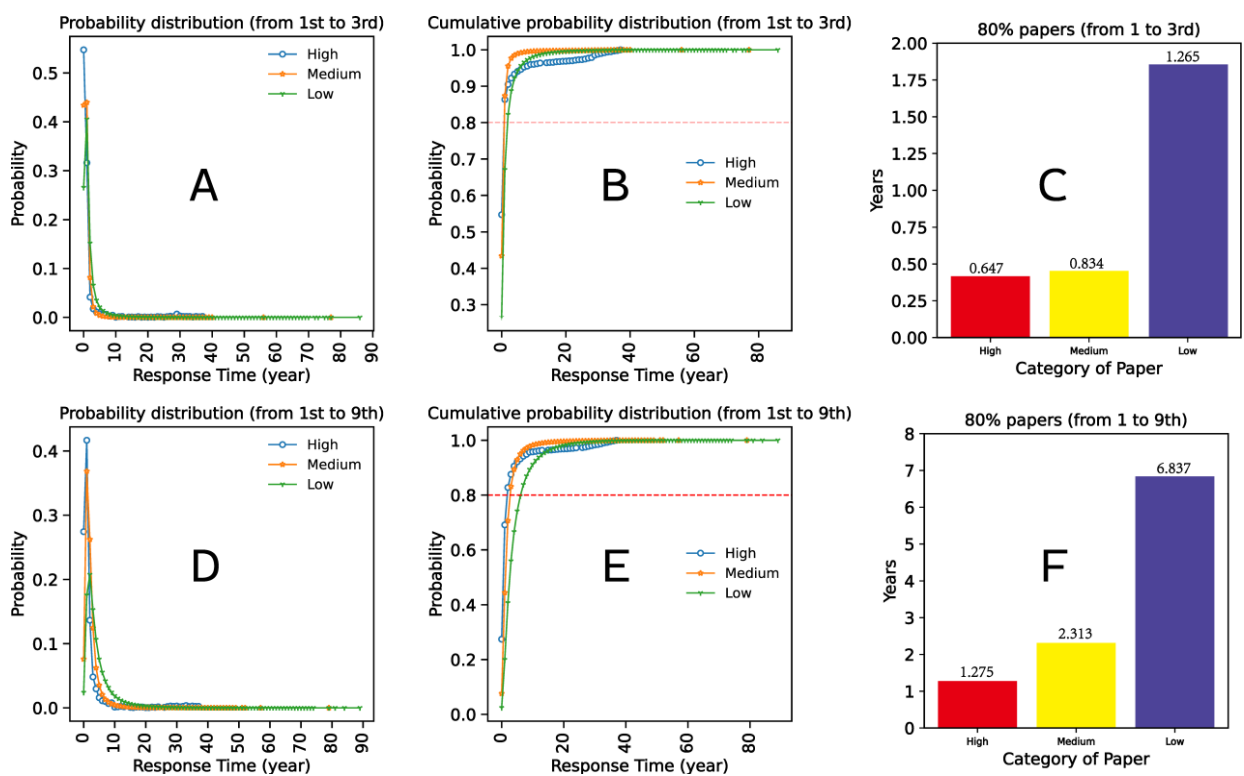

Fig.S3 Probability distributions of response time for biomedical papers with different clinical citation levels (i.e., high, medium, and low) to receive citations from general literature [from 1st to 3rd] (**A**, **B**). Comparisons of the time required for 80% of biomedical papers with different clinical citation levels to receive citations from general literature [from 1st to 3rd] (**C**). Probability distributions of response time for biomedical papers with different clinical citation levels to receive citations from general literature [from 1st to 9th] (**D**, **E**). Comparisons of the time required for 80% of biomedical papers with different clinical citation levels to receive citations from general literature [from 1st to 9th] (**F**).

**Appendix 4 The PD and PCD of response time for four categories of biomedical papers to receive their citation from general literature from $1^{st}$ to $N^{th}$.**

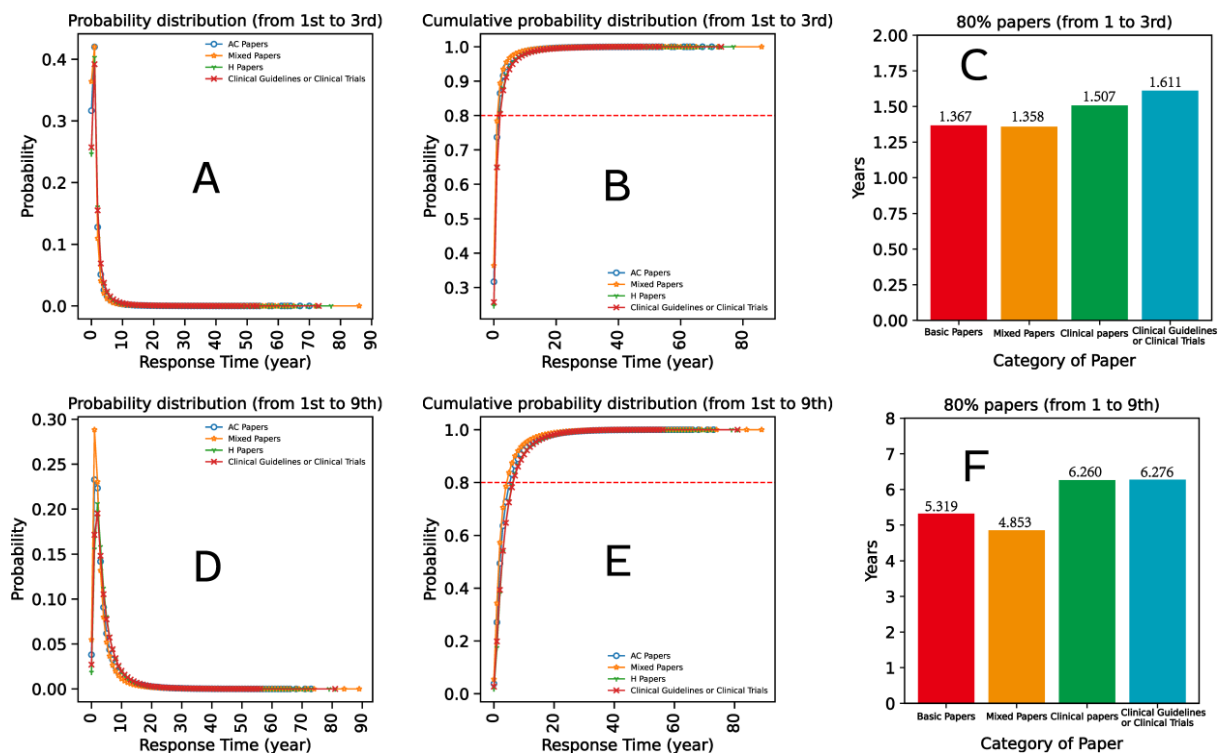

Fig. S4 Probability distributions of response time for four categories of biomedical papers to receive citations from general literature [from $1^{st}$ to $3^{rd}$] (**A**, **B**). Comparisons of the time required for 80% of biomedical papers in four categories to receive citations from general literature [from $1^{st}$ to $3^{rd}$] (**C**). Probability distributions of response time for four categories of biomedical papers to receive citations from general literature [from $1^{st}$ to $9^{th}$] (**D**, **E**). Comparisons of the time required for 80% of biomedical papers in four categories to receive citations from general literature [from $1^{st}$ to $9^{th}$] (**F**).